\documentclass[pra,aps,amsmath,amssymb,amsfonts,twocolumn,nofootinbib,longbibliography]{revtex4-1}
\usepackage{amssymb}
\usepackage{bm,mathrsfs}
\usepackage{graphicx}
\usepackage{epsfig}
\usepackage{amsmath,bbm}
\usepackage{amsfonts,amssymb}
\usepackage{times}
\usepackage{verbatim}
\usepackage[sort&compress]{natbib}
\usepackage{amsmath}
\usepackage[colorlinks=true,citecolor=blue,linkcolor=blue,urlcolor=blue]{hyperref}
\usepackage[usenames]{color}

\newcommand{\bra}[1]{\langle #1|}
\newcommand{\ket}[1]{|#1\rangle}

\begin{document}

\title{Dicke States Generation via Selective Interactions in Dicke-Stark Model}

\author{Fengchun Mu}
\affiliation{Center for Quantum Sciences and School of Physics, Northeast Normal University, Changchun 130024, China}

\author{Ya Gao}
\affiliation{Center for Quantum Sciences and School of Physics, Northeast Normal University, Changchun 130024,  China}

\author{Hong-Da Yin}
\affiliation{Center for Quantum Sciences and School of Physics, Northeast Normal University, Changchun 130024, China}

\author{Gangcheng Wang}
\email{wanggc887@nenu.edu.cn}
\affiliation{Center for Quantum Sciences and School of Physics, Northeast Normal University, Changchun 130024, China}

\date{\today}

\begin{abstract}
We propose a method to create selective interactions with Dicke-Stark model by means of time-dependent perturbation theory. By choosing the proper rotating framework, we find that the time oscillating terms depend on the number of atomic excitations and the number of photonic excitations. Consequently, the Rabi oscillation between selective states can be realized by properly choosing the frequency of the two-level system. The second order selective interactions can also be studied with this method. Then various states such as Dicke states, superposition of Dicke states and GHZ states can be created by means of such selective interactions. The numerical results show that high fidelity Dicke states and Greenberger-Horne-Zeilinger states can be created by choosing the proper frequency of two-level system and controlling the evolution time.
\end{abstract}

\maketitle

\section{Introduction}
\label{SecI}
Quantum entanglement is one of the most prominent properties of quantum states that has no classical analog \cite{RMP2009Horodecki}. So far, a variety types entanglement states have been studied theoretically and experimentally \cite{GHZ1989,RMP2012Pan,PRA2000Dur,NP2009Briegel}. It has also motivated a variety of quantum protocols in quantum information theory \cite{Nielsen2010}, such as teleportation \cite{PRL1993Bennett}, dense coding \cite{PRL1992Bennett}, and quantum key distribution \cite{PRL1991Ekert}. An important type of entangled states among them is the highly entangled Dicke states, which was firstly introduced in 1954 by Dicke \cite{PR1954Dicke}. Up to now, a growing number of efforts have been devoted to generate Dicke states \cite{PRA2004Stockton,PRA2019Masson,PRA2017Wu,PRL2009Prevedel,EPL2010Xiao}, as well as superposition Dicke states, in a wide variety of quantum platforms, such as trapped ions \cite{PRA1994Cirac,PRL2003Unanyan} and circuit QED system \cite{PRA2018Kasture,PRL2002Hong,JPB2005Xiao,JOB2003Yu,AOP2018Ran}. 

On the other hand, selective interactions have a wide range of applications in quantum information theory, such as entanglement states generation. Such special interactions arise from the possibility of tuning to resonance transitions inside a chosen Hilbert subspace, while leaving other transitions dispersive \cite{MC2001Solano}. In \cite{PRA2000Solano}, Solano et al proposed the selective interactions permits to manipulate motional states in two trapped ions. Recently, in ~\cite{PRA2020Cong}, Cong et al studied the selective $k$-photonic interactions in the quantum Rabi model with Stark term, which is termed Rabi-Stark model \cite{PRA2013Grimsmo,PRA2014Grimsmo,JPA2017Hans} and selective interactions can be used to create photonic Fock states. Such model attract much attentions in recent years \cite{PLA2015Maciejewski, JPA2005Xie, CTP2019Xie}. The Stark term plays an important role in dynamical selectivity. This nonlinear coupling term was also added to the Dicke model \cite{PTRSA2011Garraway, PRL2011Gopalakrishnan,PRL2012Bastidas,IJM2017Abdel}, which is a fundamental model of quantum optics to describe the interactions between light and matter \cite{AQT2019Kirton,PRA2012Bhaseen,PRA2007Dimer,IOP2013Grimsmo,PRA2018Zhang}. The existence of Stark term in the Dicke-Stark (DS) model can be used to create nonlinear energy levels, which will be used to generate entangled states selectively. 

In this work, we will study the selective interactions in the DS model. Choosing proper rotating frame, we find the time-varying terms depend on the number of excitations. Then one can choose proper frequencies of the system to obtain selective Tavis-Cummings (TC) \cite{PR1968Tavis} or anti-TC interactions from the DS model \cite{Nature2009Gunter,NP2010Niemczyk,PRA2017Rossatto,NRP2019Frisk,RMP2019Forn}. Considering the second order effects, the two-atom selective interactions also can be obtained. Acting such effective interactions on the pre-selective initial states, one can obtain the selective target Dicke states, as well as superposition of Dicke states. Finally, the validity of our proposal is studied with numerical simulations. The results show that one can obtain high fidelity by properly choosing frequencies of two-level systems.


\section{The derivation of the effective Hamiltonian}

\label{SecII}
The Hamiltonian of DS model, describing the Dicke model with Stark term, is given by ($\hbar = 1$)
\begin{equation}\label{eq1}
  \hat{H}_{\rm DS} = \hat{H}_{0} + \hat{H}_{\rm int},
\end{equation}
where
\begin{equation*}
\begin{split}
  \hat{H}_{0} &= \frac{\omega_{q}}{2}\sum_{j=1}^{N}\hat{\sigma}_{j}^{z}+\omega_{r}\hat{a}^{\dagger}\hat{a}
+\frac{U}{2N}\hat{a}^{\dagger}\hat{a}\sum_{j=1}^{N}\hat{\sigma}_{j}^{z},\\
  \hat{H}_{\rm int} &= \frac{\lambda}{\sqrt{N}}\sum_{j=1}^{N}(\hat{a}+\hat{a}^{\dagger})\hat{\sigma}_{j}^{x}.
\end{split}
\end{equation*}
This Hamiltonian describes $N$ two-level systems (or qubits) with uniform transition frequency $\omega_{q}$ coupling to a single mode bosonic field with frequency $\omega_{r}$. Here we assume the coupling strength between each qubit and resonator is uniform. The parameter $U$ is the coupling strength of the Stark term. $\hat{a}$ and $\hat{a}^{\dagger}$ denote the annihilation and creation operators of the resonator respectively. The $j$-th qubit operators are $\hat{\sigma}_{j}^{x}=\ket{e_{j}} \bra{g_{j}}+\ket{g_{j}}\bra{e_{j}}$, $\hat{\sigma}_{j}^{y}=-i(\ket{e_{j}}\bra{ g_{j}}-\ket{g_{j}}\bra{e_{j}})$ and $\hat{\sigma}_{j}^{z}=\ket{e_{j}}\bra{e_{j}}-\ket{g_{j}}\bra{g_{j}}$ with $\ket{g_{j}}$ and $\ket{e_{j}}$ be the ground and excited states for the $j$-th qubit. By utilizing the relation $\hat{\sigma}_{j}^{\pm}=(\hat{\sigma}_{j}^{x}\pm i\hat{\sigma}_{j}^{y})/2$, the Dicke coupling term in the Hamiltonian Eq.~(\ref{eq1}) can be rewritten as following two terms $(\lambda/\sqrt{N})\sum_{j=1}^{N}(\hat{a}\hat{\sigma}_{j}^{+}+\hat{a}^{\dagger}\hat{\sigma}_{j}^{-})$ and $(\lambda/\sqrt{N})\sum_{j=1}^{N}(\hat{a}^{\dagger}\hat{\sigma}_{j}^{+}+\hat{a}\hat{\sigma}_{j}^{-})$, which corresponding to rotating and counter-rotating terms. In the situation where the qubits are near resonance with the resonator and the coupling between qubits and resonator are much smaller than the transition frequency of qubits and frequency of the resonator, the rotating-wave-approximation (RWA) is valid. In this case, we can drop the counter-rotating term safely. Consequently, the Hamiltonian in Eq.~(\ref{eq1}) reduced to the Tavis-Cummings (TC) model with Stark term. The Dicke model with Stark term in Eq.~(\ref{eq1}) depends on the following atomic collective operators $\hat{J}_{\alpha} = \sum_{j=1}^{N}\hat{\sigma}_{j}^{\alpha}$ with $\alpha=x,z$. In terms of atomic collective operators, the Hamiltonian in Eq.~(\ref{eq1}) reads
\begin{equation}
\begin{split}
\hat{H}_{\rm DS}&=\frac{\omega_{q}}{2}\hat{J}_{z}+\omega_{r}\hat{a}^{\dagger}\hat{a}
+\frac{\lambda}{\sqrt{N}}(\hat{a}+\hat{a}^{\dagger})\hat{J}_{x}
+\frac{U}{2N}\hat{a}^{\dagger}\hat{a}\hat{J}_{z}
.\\
\label{eq1-2}
\end{split}
\end{equation}
To obtain a closed analytical description, we introduce the normalized Dicke states with $k$ atomic excitations \cite{PRL2012Noguchi,PRA2011Zhou,PRA2009Hume}
\begin{equation}
\left|D_{N}^{k}\right\rangle \equiv\left(C^{N}_{k}\right)^{-1 / 2} \sum_{m=1}^{C^{N}_{k}} P_{m}\ket{\overbrace{e_{j_{1}},e_{j_{2}}\cdots,e_{j_{k}}}^{k},\overbrace{g_{j_{k+1}},\cdots,g_{j_{N}}}^{N-k}}.
\label{eq2}
\end{equation}
Here $\sum_{m=1}^{C^{N}_{k}} P_{m}(\cdots)$ indicates the sum over all particle permutations and $C^{N}_{k}=N!/[k!(N-k)!]$. In the Dicke states basis, the collective operators can be reduced to $\hat{J}_{x,z}^{D}$ as
\begin{equation}\label{eq2-2}
\begin{split}
  \hat{J}_{x}^{D} &= \sum_{k=0}^{N}f(k)(\ket{D_{N}^{k+1}}\bra{D_{N}^{k}}+\ket{D_{N}^{k}}\bra{D_{N}^{k+1}}), \\
  \hat{J}_{z}^{D} &= \sum_{k=0}^{N}(2k-N)\ket{D_{N}^{k}}\bra{D_{N}^{k}},
  \end{split}
\end{equation}
where $f(k)=\sqrt{(k+1)(N-k)}$. In terms of Dicke states and Fock states, the Hamiltonian Eq.~(\ref{eq1}) can be recast as follows
\begin{equation}
\begin{split}
  &\hat{H}^{D}_{0} = \sum_{n,k}\left[\left(\omega_{q}+n\frac{U}{N}\right)\left(k-\frac{N}{2}\right)+n\omega_{r}
\right] \hat{D}_{k,k}\otimes\hat{A}_{n,n}, \\
  &\hat{H}^{D}_{\rm int} = \sum_{n,k} \Omega_{nk}(\lambda)(\hat{D}_{k+1,k}+\hat{D}_{k,k+1})\otimes(\hat{A}_{n,n+1}+\hat{A}_{n+1,n}),
  \end{split}
\end{equation}
where $\Omega_{nk}(\lambda)=\lambda f(k)\sqrt{n+1}/\sqrt{N}$, $\hat{D}_{k,k'}=\ket{D_{N}^{k}}\bra{D_{N}^{k'}}$, and $\hat{A}_{n,n'}=\ket{n}\bra{n'}$. Moving to the interaction picture with respect to $\hat{H}_{0}^{D}$, one obtains the following transformed Hamiltonian
\begin{equation}
\begin{split}
\hat{H}_{I}^{D}(t)&=\hat{R}^{\dagger}(t)\left(\hat{H}_{0}^{D}+\hat{H}_{\rm int}^{D}\right)\hat{R}(t)+i\left[\partial_{t}\hat{R}^{\dagger}(t)\right]\hat{R}(t)\\
                        &=\hat{H}_{\rm TC}^{D}(t)+\hat{H}_{\rm aTC}^{D}(t),
\label{eq4}
\end{split}
\end{equation}
where $\hat{R}(t)=\exp(-i\hat{H}_{0}^{D}t)$. The TC and anti-TC Hamiltonian in the rotating framework are as follows
\begin{equation}\label{HD_I}
\begin{split}
  &\hat{H}_{\rm TC}^{D}(t)=\sum_{n,k}\Omega_{nk}(\lambda)\left(\hat{D}_{k+1,k}\otimes\hat{A}_{n,n+1}e^{i\delta_{nk}^{-}t}+{\rm H.c.}\right),\\
 & \hat{H}_{\rm aTC}^{D}(t)=\sum_{n,k}\Omega_{nk}(\lambda)\left(\hat{D}_{k+1,k}\otimes\hat{A}_{n+1,n}e^{i\delta_{nk}^{+}t}+{\rm H.c.}\right).
  \end{split}
\end{equation}
Here $\delta_{nk}^{+}=\omega_{r}+\omega_{q}+U(n+k+1-N/2)/N$, $\delta_{nk}^{-}=\omega_{q}-\omega_{r}+U(n-k+N/2)/N$, and H.c. denotes the Hermitian conjugate. Obviously, the resonance frequencies $\delta_{nk}^{\pm}$ depend on the photon number $n$ and the atomic excitation number $k$. We can tune the parameters to obtain the resonant transition, and the other transitions are off-resonance. If we fix the photon number $n=n_{0}$ in the initial states, the oscillation frequencies $\delta_{nk}^{\pm}$ only depend on $k$. Consequently, the designed interactions depend on $k$ can be realized. If $U=0$, the detunings $\delta_{nk}^{\pm}$ are $n$ and $k$ independent. For $|\delta^{+}|\gg |\Omega_{nk}(\lambda)|$ and $\delta^{-}=0$, $\hat{H}^{D}_{\rm TC}$ is recovered when fast oscillating terms are averaged out by utilizing the RWA. Under these conditions, the dynamics lead to Rabi oscillations between the states $\ket{D^{k+1}_{N}}\otimes \ket{n}\leftrightarrow \ket{D^{k}_{N}}\otimes \ket{n+1}$ for each pair of $n$ and $k$ with the Rabi frequency $\Omega_{nk}(\lambda)$. On the contrary, the conditions $|\delta^{-}|\gg |\Omega_{nk}(\lambda)|$ and $\delta^{+}=0$ lead to an anti-TC resonant interaction. Consequently, the dynamics lead to Rabi oscillations between states $\ket{D^{k}_{N}}\otimes \ket{n}$ and $\ket{D^{k+1}_{N}}\otimes \ket{n+1}$. The interactions realized in such case are not selective as they apply to all pairs of $n$ and $k$.

\begin{figure}[ht]
\includegraphics[angle=0,width=0.40\textwidth]{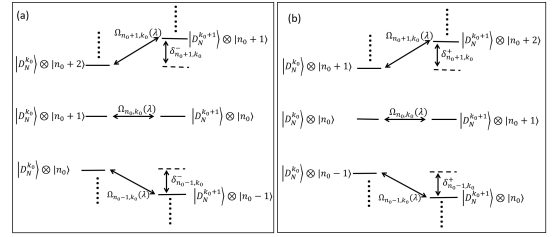}
\caption{The sketches of energy level for the selected TC and anti-TC model: (a) shows that only selected states $\ket{D_{N}^{k_{0}}}\otimes \ket{n_{0}+1}$ and $\ket{D_{N}^{k_{0}+1}}\otimes \ket{n_{0}}$ are resonant and they are detuning with the other states; (b) shows that only selected states $\ket{D_{N}^{k_{0}}}\otimes \ket{n_{0}}$ and $\ket{D_{N}^{k_{0}+1}}\otimes \ket{n_{0}+1}$ are resonant and they are detuning with the other states.}
\label{fig1}
\end{figure}

The presence of a non-zero Stark coupling $U$ makes the TC and anti-TC resonant interactions different for the Fock state labeled by $n$ and Dicke state labeled by $k$. Given different values of $n$ and $k$, it is possible to adjust the value of the parameters to identify a resonance condition that applies only for a selected state. We note that if $\delta_{n_{0}+m,k_{0}+m}^{-}=0$, $|\delta_{n\neq n_{0}+m, k\neq k_{0}+m}^{-}|\gg|\Omega_{n\neq n_{0}+m, k\neq k_{0}+m}(\lambda)|$ and $|\delta_{nk}^{+}|\gg|\Omega_{nk}(\lambda)|$, the dynamics of Hamiltonian Eq.~(\ref{eq4}) will produce a selective TC interaction, which leads to transitions between $\ket{D^{k_{0}+m+1}_{N}}\otimes \ket{n_{0}+m}$ and $\ket{D^{k_{0}+m}_{N}}\otimes \ket{n_{0}+m+1}$. If the initial state is prepared to the fix photon number $n= n_{0}$ (i.e., $m=0$), only interaction between $\ket{D^{k_{0}+1}_{N}}\otimes \ket{n_{0}}$ and $\ket{D^{k_{0}}_{N}}\otimes \ket{n_{0}+1}$ is survived. The effective Hamiltonian for such selective TC interaction reads
\begin{equation}\label{HTC}
  \hat{H}_{\rm TC}^{\rm eff}=\Omega_{n_{0},k_{0}}(\lambda)\left(\hat{D}_{k_{0}+1,k_{0}}\otimes\hat{A}_{n_{0},n_{0}+1}+{\rm H.c.}\right).
\end{equation}
The case in which $\delta_{n_{0}+m,k_{0}-m}^{+}=0$, $|\delta_{n\neq n_{0}+m, k\neq k_{0}-m}^{+}|\gg|\Omega_{n\neq n_{0}+m, k\neq k_{0}-m}(\lambda)|$ and $\delta_{nk}^{-}\gg\Omega_{nk}(\lambda)$ leads to transitions between $\ket{D^{k_{0}-m}_{N}}\otimes \ket{n_{0}+m}$ and $\ket{D^{k_{0}-m+1}_{N}}\otimes \ket{n_{0}+m+1}$. If the initial state is prepared to the fix photon number $n= n_{0}$ (i.e., $m=0$), only interaction between $\ket{D^{k_{0}}_{N}}\otimes \ket{n_{0}}$ and $\ket{D^{k_{0}+1}_{N}}\otimes \ket{n_{0}+1}$ is survived. The effective Hamiltonian for the anti-TC interaction reads
\begin{equation}\label{HaTC}
  \hat{H}_{\rm aTC}^{\rm eff}=\Omega_{n_{0},k_{0}}(\lambda)\left(\hat{D}_{k_{0}+1,k_{0}}\otimes\hat{A}_{n_{0}+1,n_{0}}+{\rm H.c.}\right).
\end{equation}
The sketches of energy level for the selected TC and anti-TC model are shown in Fig. \ref{fig1}. The Fig. \ref{fig1}(a) shows that only selected states $\ket{D_{N}^{k_{0}}}\otimes \ket{n_{0}+1}$ and $\ket{D_{N}^{k_{0}+1}}\otimes \ket{n_{0}}$ are resonant and they are detuning with the other states. The effective Hamiltonian describing such selected TC interaction are given in Eq. (\ref{HTC}). The Fig. \ref{fig1}(b) shows that selected states $\ket{D_{N}^{k_{0}}}\otimes \ket{n_{0}}$ and $\ket{D_{N}^{k_{0}+1}}\otimes \ket{n_{0}+1}$ are resonant and they are detuning with the other states.

\begin{figure}[ht]
\centering
\includegraphics[angle=0,width=0.45\textwidth]{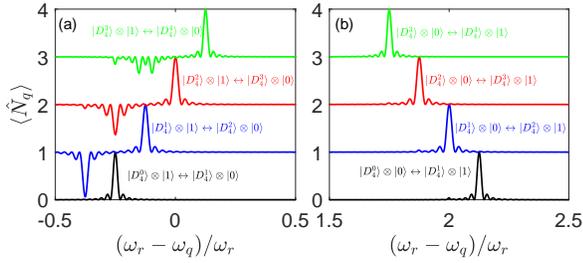}
\caption{The selection interaction in Dicke-Stark model: (a) shows the resonance transition between $\ket{D_{4}^{k_{0}}}\otimes \ket{1}$ and $\ket{D_{4}^{k_{0}+1}}\otimes \ket{0}$, which controlled by the selected TC model in Eq.~(\ref{HTC}); (b) shows the resonance transition between $\ket{D_{4}^{k_{0}}}\otimes \ket{0}$ and $\ket{D_{4}^{k_{0}+1}}\otimes \ket{1}$, which controlled by the selected anti-TC model in Eq.~(\ref{HaTC}). Other parameters are given 
as follows: $\omega_{r}=1$, $\lambda/\omega_{r}=0.006$ and $U/ \omega_{r}=-0.5$. Here $\hat{N}_{q} = \sum_{j}^{N}\hat{\sigma}_{j}^{+}\hat{\sigma}_{j}^{-}$ denotes the atomic excitation operator.}
\label{fig2}
\end{figure}

The selective resonance peaks for initial states with different number of atomic excitations are shown in Fig. \ref{fig2}, where four two-level systems are considered. The Fig. \ref{fig2}(a) shows the Hamiltonian acts on the initial states $\ket{D_{4}^{k_{0}}}\otimes \ket{1}$ during a time $t=\pi/(2\Omega_{n_{0},k_{0}}(\lambda))$. The results show that only the resonance transition between $\ket{D_{4}^{k_{0}}}\otimes \ket{1}$ and $\ket{D_{4}^{k_{0}+1}}\otimes \ket{0}$ can occur. The Fig. \ref{fig2}(b) shows the Hamiltonian acts on the initial state $\ket{D_{4}^{k_{0}}}\otimes \ket{0}$, and only the resonance transition between $\ket{D_{4}^{k_{0}}}\otimes \ket{0}$ and $\ket{D_{4}^{k_{0}+1}}\otimes \ket{1}$ can occur. The average atomic excitation number $\langle \hat{N}_{q}\rangle$ also shows the oscillatory behavior near the resonance peaks. These behaviors can be explained by Rabi oscillating with small detuning. We take the transition $\ket{D_{4}^{0}}\otimes\ket{1}\leftrightarrow \ket{D_{4}^{1}}\otimes\ket{0}$ as an example (the solid black line in left panel of fig. \ref{fig2}). There is a resonance peak locate at $(\omega_{r}-\omega_{q})/\omega_{r}=-0.250$. If we consider there is a small detuning $\delta$, the effective Hamiltonian in Eq. (\ref{HTC}) can be recast as follows
\begin{equation}\label{xHTC}
\hat{H}_{\rm TC}^{\rm eff}=\Omega_{1,0}(\lambda)\left(\hat{D}_{1,0}\otimes\hat{A}_{0,1}e^{i\delta t}+{\rm H.c.}\right).
\end{equation}
Then the probability of the system in state $\ket{D_{4}^{1}}\otimes\ket{0}$ at time $t$ can be obtained as follows
\begin{equation}
P_{\ket{D_{4}^{1}}\otimes\ket{0}}(t)= \frac{4(\Omega_{1,0}(\lambda))^{2}}{4(\Omega_{1,0}(\lambda))^{2}+\delta^{2}}\sin^{2}\left(\frac{1}{2}\sqrt{4(\Omega_{1,0}(\lambda))^{2}+\delta^{2}}t\right).
\end{equation}
At $t_{0}=\pi/(2\Omega_{1,0}(\lambda))$, we obtain the probabilities of the system in the state $\ket{D_{4}^{1}}\otimes\ket{0}$ is
\begin{equation}
P_{\ket{D_{4}^{1}}\otimes\ket{0}}(t_{0})= \frac{4(\Omega_{1,0}(\lambda))^{2}}{4(\Omega_{1,0}(\lambda))^{2}+\delta^{2}}\sin^{2}\left(\frac{t_{0}}{2}\sqrt{4(\Omega_{1,0}(\lambda))^{2}+\delta^{2}}\right).
\end{equation}
Near the resonance peak, the probability and $\langle \hat{N}_{q}\rangle$ show oscillatory behavior. When the detuning $\delta=0$, the probabilities of the system in the state $\ket{D_{4}^{1}}\otimes\ket{0}$ will reach its maximum value $1$. With the increases of detuning, the amplitude of the oscillations is attenuated. Such similar behavior also is studied in \cite{PRA2017Huang}.

The DS model not only can be used to produce single-atom excitation interactions, but also can be used to generate more than single-atom excitation interactions. To show multi-atom excitation process, we should consider higher order effective Hamiltonian. Here, we will consider two-atom excitation process. By means of the effective Hamiltonian method given in \cite{PRA2017shao}, we will derive the second order effective Hamiltonian. If all terms in the Eq.~(\ref{HD_I}) are fast-time varying terms, higher order effective Hamiltonian will dominate the dynamics. According to \cite{CJP2007James}, the second order Hamiltonian can be derived as follows 
\begin{widetext}
\begin{equation}
\begin{split}
\hat{H}^{(2)}(t) &=\hat{H}_{0}^{(2)}+\hat{H}_{\rm TC}^{(2)}(t)+\hat{H}_{\rm aTC}^{(2)}(t)+\hat{H}_{r}^{(2)}(t)+\hat{H}_{a}^{(2)}(t),
\label{eq_o2_H}
\end{split}
\end{equation}
where

\begin{equation}
\begin{split}
& \hat{H}_{0}^{(2)}=\sum_{n,k}\Delta_{n,k}\hat{D}_{k,k}\otimes \hat{A}_{n,n},\\
& \hat{H}_{\rm TC}^{(2)}(t)=\sum_{n,k}\Omega_{\rm TC}^{(2)}\left(\hat{D}_{k+2,k}\otimes\hat{A}_{n,n+2}e^{i\delta_{TC}^{(2)}(n,k)t}+{\rm H.c.}\right),\\
& \hat{H}_{\rm aTC}^{(2)}(t)=\sum_{n,k}\Omega_{\rm aTC}^{(2)}\left(\hat{D}_{k+2,k}\otimes\hat{A}_{n+2,n}e^{i\delta_{aTC}^{(2)}(n,k)t}+{\rm H.c.}\right),\\
& \hat{H}_{ r}^{(2)}(t)=\sum_{n,k}\Omega_{r}^{(2)}\left(\hat{D}_{k+2,k}\otimes \hat{A}_{n,n}e^{i\delta_{r}^{(2)}(n)t}+{\rm H.c.}\right),\\
& \hat{H}_{a}^{(2)}(t)=\sum_{n,k}\Omega_{a}^{(2)}\left(\hat{D}_{k,k}\otimes \hat{A}_{n+2,n}e^{i\delta_{a}^{(2)}(k)t}+{\rm H.c.}\right).
\end{split}
\end{equation}
Here 

\begin{equation}
\begin{split}
& \Delta_{n,k}=\frac{\Omega^{2}_{n,k-1}(\lambda)}{\delta^{-}_{n,k-1}}+\frac{\Omega^{2}_{n-1,k-1}(\lambda)}{\delta^{+}_{n-1,k-1}}-\frac{\Omega^{2}_{n-1,k}(\lambda)}{\delta^{-}_{n-1,k}}-\frac{\Omega^{2}_{n,k}(\lambda)}{\delta^{+}_{n,k}},\\
& \Omega_{\rm TC}^{(2)}=\frac{1}{2}\Omega_{n,k+1}(\lambda)\Omega_{n+1,k}(\lambda)\left(\frac{1}{\delta_{n,k+1}^{-}}-\frac{1}{\delta_{n+1,k}^{-}}\right),\\
& \Omega_{\rm aTC}^{(2)}=\frac{1}{2}\Omega_{n,k}(\lambda)\Omega_{n+1,k+1}(\lambda)\left(\frac{1}{\delta_{n+1,k+1}^{+}}-\frac{1}{\delta_{n,k}^{+}}\right),\\
& \Omega_{r}^{(2)}=\frac{1}{2}\left(\Omega_{n-1,k}(\lambda)\Omega_{n-1,k+1}(\lambda)\left(\frac{1}{\delta_{n-1,k+1}^{+}}-\frac{1}{\delta_{n-1,k}^{-}}\right)+\Omega_{n,k}(\lambda)\Omega_{n,k+1}(\lambda)\left(\frac{1}{\delta_{n,k+1}^{-}}-\frac{1}{\delta_{n,k}^{+}}\right)\right),\\
& \Omega_{a}^{(2)}=\frac{1}{2}\left(\Omega_{n,k-1}(\lambda)\Omega_{n+1,k-1}(\lambda)\left(\frac{1}{\delta_{n+1,k-1}^{+}}+\frac{1}{\delta_{n,k-1}^{-}}\right)-\Omega_{n,k}(\lambda)\Omega_{n+1,k}(\lambda)\left(\frac{1}{\delta_{n,k}^{+}}+\frac{1}{\delta_{n+1,k}^{-}}\right)\right),
\end{split}
\end{equation}
\end{widetext}

and $\delta_{\rm TC}^{(2)}(n,k)=\delta_{n+1,k}^{-}+\delta_{n,k+1}^{-}$, $\delta_{\rm aTC}^{(2)}(n,k)=\delta_{n,k}^{+}+\delta_{n+1,k+1}^{+}$, $\delta_{r}^{(2)}(n,k)=\delta_{n,k}^{+}+\delta_{n,k+1}^{-}$, $\delta_{a}^{(2)}(n,k)=\delta_{n,k}^{+}-\delta_{n+1,k}^{-}$. The total excitation number conserved term $\hat{H}_{0}^{(2)}$ in Hamiltonian Eq.~(\ref{eq_o2_H}) is effective Stark shift. Moving to the rotating frame with respect to $\hat{R}_{s}(t)=e^{-i \hat{H}_{0}^{(2)}t}$, we obtain the following transformed Hamiltonian 
\begin{equation}
    \tilde{H}^{(2)}(t) = \tilde{H}^{(2)}_{\rm TC}(t)+\tilde{H}^{(2)}_{\rm aTC}(t)+\tilde{H}^{(2)}_{r}(t)+\tilde{H}^{(2)}_{a}(t),
\end{equation}
where $\hat{\tilde{H}}^{(2)}_{\alpha}(t)=\hat{R}_{s}^{\dagger}(t)\hat{H}_{\alpha}^{(2)}\hat{R}_{s}(t)+i\left[\partial_{t}\hat{R}_{s}^{\dagger}(t)\right]\hat{R}_{s}(t)$ with $\alpha = {\rm TC},~{\rm aTC},~r,~a$. Then the time oscillating frequency can be modified to $\tilde{\delta}_{\rm TC}^{(2)}(n,k)=\delta_{\rm TC}^{(2)}(n,k)+\Delta_{n,k+2}-\Delta_{n+2,k}$, $\tilde{\delta}_{\rm aTC}^{(2)}(n,k)=\delta_{\rm aTC}^{(2)}(n,k)+\Delta_{n+2,k+2}-\Delta_{n,k}$, $\tilde{\delta}_{r}^{(2)}(n,k)=\delta_{r}^{(2)}(n,k)+\Delta_{n,k+2}-\Delta_{n,k}$ and $\tilde{\delta}_{a}^{(2)}(n,k)=\delta_{a}^{(2)}(n,k)+\Delta_{n+2,k}-\Delta_{n,k}$. In the following, we will focus on the two-atom TC and anti-TC dynamics. Obviously, the time oscillating frequencies $\tilde{\delta}_{\rm TC}^{(2)}(n,k)$, $\tilde{\delta}_{\rm aTC}^{(2)}(n,k)$ depend on $n$ and $k$. In order to get an effective TC and anti-TC Hamiltonian, we can choose proper frequencies of two-level systems to make $\tilde{\delta}_{\rm TC}^{(2)}(n,k)=0$ or $\tilde{\delta}_{\rm aTC}^{(2)}(n,k)=0$ and the other time oscillating frequencies large greater than its corresponding effective coupling strength. Then the fast oscillating terms are averaged out by utilizing the RWA. By means of these conditions, we can achieve the so-called selective two photon TC and anti-TC model.

Under conditions $\tilde{\delta}_{\rm TC}^{(2)}(n_{0}+m,k_{0}+m)=0$, $|\tilde{\delta}_{\rm TC}^{(2)}(n\neq n_{0}+m,k\neq k_{0}+m)|\gg \Omega_{\rm TC}^{(2)}(n\neq n_{0}+m,k\neq k_{0}+m,\lambda) $ and $|\tilde{\delta}_{\alpha}^{(2)}(n,k)|\gg \Omega_{\alpha}^{(2)}(n,k,\lambda) $ with $\alpha= {\rm aTC}$, $r$, or $a$, the dynamic evolution leads to transitions between $\ket{D^{k_{0}+m+2}_{N}}\otimes \ket{n_{0}+m}$ and $\ket{D^{k_{0}+m}_{N}}\otimes \ket{n_{0}+m+2}$. Fixing the photon number $n=n_{0}$ (i.e. $m=0$), the interaction between $\ket{D^{k_{0}+2}_{N}}\otimes \ket{n_{0}}$ and $\ket{D^{k_{0}}_{N}}\otimes \ket{n_{0}+2}$ is survived. The effective Hamiltonian for the TC interaction reads
 \begin{equation}\label{H2TC}
\hat{H}_{\rm TC}^{(2)}=\Omega_{\rm TC}^{(2)}(n_{0},k_{0},\lambda)\left(\hat{D}_{k_{0}+2,k_{0}}\otimes\hat{A}_{n_{0},n_{0}+2}+{\rm H.c.}\right).
\end{equation}
We also note that if $\tilde{\delta}_{\rm aTC}^{(2)}(n_{0}+m,k_{0}-m)=0$, $|\tilde{\delta}_{\rm aTC}^{(2)}(n\neq n_{0}+m,k\neq k_{0}-m)|\gg \Omega_{\rm aTC}^{(2)}(n\neq n_{0}+m,k\neq k_{0}-m,\lambda)$, $|\tilde{\delta}_{\alpha}^{(2)}(n,k)|\gg \Omega_{\alpha}^{(2)}(n,k,\lambda) $ with $\alpha = {\rm TC}$, $r$, or $a$, only interaction between $\ket{D^{k_{0}}_{N}}\otimes \ket{n_{0}}$ and $\ket{D^{k_{0}+2}_{N}}\otimes \ket{n_{0}+2}$ is survived (i.e. $m=0$). The effective Hamiltonian for anti-TC interaction reads
\begin{equation}\label{H2aTC}
\hat{H}_{\rm aTC}^{(2)}=\Omega_{\rm aTC}^{(2)}(n_{0},k_{0},\lambda)\left(\hat{D}_{k_{0}+2,k_{0}}\otimes\hat{A}_{n_{0}+2,n_{0}}+{\rm H.c.}\right).
\end{equation}
The higher order dynamics can also be considered with this method. Then the higher order selective TC and anti-TC effective Hamiltonian can be achieved.

\section{The applications of selective interaction}
\label{SecIII}
In this section, we will show how to generate Dicke states and superposition of Dicke states with the selective TC and anti-TC interactions given in Sec. \ref{SecII}. Here we will introduce our method by taking $N=4$. 

\subsection{Generation Dicke states with single-atom excitation TC and anti-TC model}
\begin{figure}[ht]
\centering
\includegraphics[angle=0,width=0.45\textwidth]{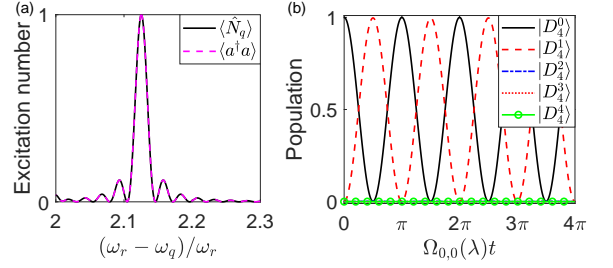}
\caption{One-atom-excitation selective interactions of the Dicke-stark model. (a) After a time $T_{1}=\pi/(2\Omega_{0,0}(\lambda))$, we calculation the average number of excitations for different ratios of $(\omega_{r}-\omega_{q})/\omega_{r}$ with initial state $\ket{D_{4}^{0}}\otimes \ket{0}$, where the anti-TC peaks appears for $(\omega_{r}-\omega_{q})/\omega_{r}=2.125$ which corresponds to $\delta_{n_{0}=0,k_{0}=0}^{+}=0$; (b) 
The population of Dicke states as function of evolution time with initial state $\ket{D_{4}^{0}}$, and the resonance frequency locate at $(\omega_{r}-\omega_{q})/\omega_{r}=2.125$. The other parameters are given as follows: $\omega_{r}=1$, $\lambda/\omega_{r}=0.006$ and $U/ \omega_{r}=-0.5$.}
\label{fig3}
\end{figure}

\begin{figure}[ht]
\centering
\includegraphics[angle=0,width=0.45\textwidth]{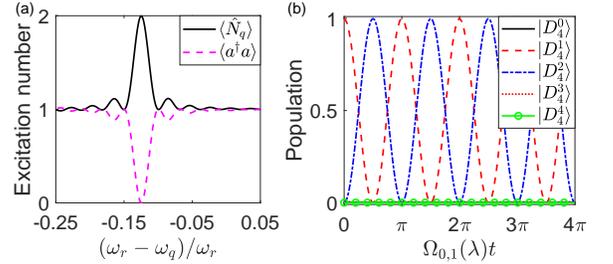}
\caption{One-atom-excitation selective interactions of the Dicke-stark model with initial state $\ket{D_{4}^{1}}\otimes \ket{1}$. (a) After a time period $T_{2}=\pi/(2\Omega_{0,1}(\lambda))$, we plot the average number of excitations for different ratios of $(\omega_{r}-\omega_{q})/\omega_{r}$ with initial state $\ket{D_{4}^{1}}\otimes \ket{1}$, where the TC transition peaks appear for $(\omega_{r}-\omega_{q})/\omega_{r}=-0.125$ which corresponds to $\delta_{n_{0}=0,k_{0}=1}^{-}=0$; (b) The population of Dicke states with initial state $\ket{D_{4}^{1}}$, and resonance frequency locate at $(\omega_{r}-\omega_{q})/\omega_{r}=-0.125$. The other parameters are given as follows: $\omega_{r}=1$, $\lambda/\omega_{r}=0.006$ and $U/ \omega_{r}=-0.5$.}
\label{fig4}
\end{figure}

\begin{figure}[ht]
\centering
\includegraphics[angle=0,width=0.45\textwidth]{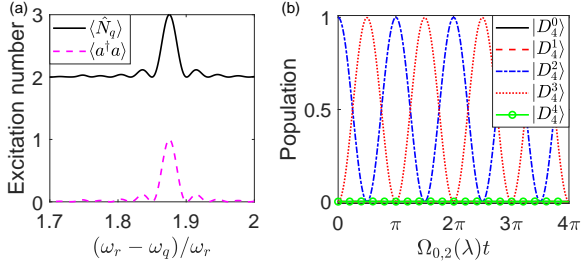}
\caption{One-atom-excitation selective interactions of the Dicke-stark model with initial state $\ket{D_{4}^{2}}\otimes \ket{0}$. (a) After a time period $T_{3}=\pi/(2\Omega_{0,2}(\lambda))$, we plot the average excitation of the particles for different ratios of $(\omega_{r}-\omega_{q})/\omega_{r}$, where the anti-TC peaks appear at $(\omega_{r}-\omega_{q})/\omega_{r}=1.875$ which corresponds to $\delta_{n_{0}=0,k_{0}=2}^{+}=0$; (b) The population of Dicke states as a function of evolution time. The resonance frequency locate at $(\omega_{r}-\omega_{q})/\omega_{r}=1.875$. The other parameters are given as follows: $\omega_{r}=1$, $\lambda/\omega_{r}=0.006$ and $U/ \omega_{r}=-0.5$.}
\label{fig5}
\end{figure}

\begin{figure}[ht]
\centering
\includegraphics[angle=0,width=0.45\textwidth]{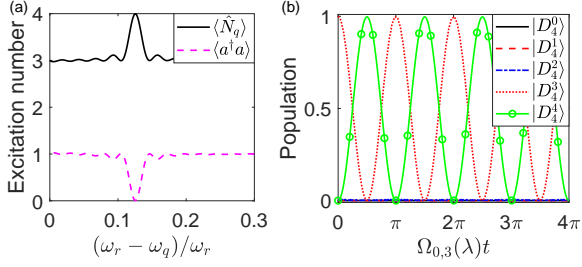}
\caption{One-atom-excitation selective interactions of the DS model with initial states $\ket{D_{4}^{3}}\otimes \ket{1}$. (a) After a time $T_{4}=\pi/(2\Omega_{0,3}(\lambda))$, we calculation the average excitation of the particles for different ratios of $(\omega_{r}-\omega_{q})/\omega_{r}$ and initial state $\ket{D_{4}^{3}}\otimes \ket{1}$, where the TC peaks appears for $(\omega_{r}-\omega_{q})/\omega_{r}=0.125$ which corresponds to $\delta_{n_{0}=0,k_{0}=3}^{-}=0$; (b) The population of Dicke states with initial state $\ket{D_{4}^{3}}$, and resonance frequency locate at $(\omega_{r}-\omega_{q})/\omega_{r}=0.125$. The other parameters are given as follows: $\omega_{r}=1$, $\lambda/\omega_{r}=0.006$ and $U/ \omega_{r}=-0.5$.}
\label{fig6}
\end{figure}

Let initial state be $\ket{\psi(0)}=\ket{D_{4}^{0}}\otimes \ket{0}\equiv\ket{g_{1}g_{2}g_{3}g_{4}}\otimes \ket{0}$. We will show how to generate Dicke states with first order Hamiltonian given in Eqs.~(\ref{HTC}) and ~(\ref{HaTC}) in several steps. First, one can tune the frequency of the two-level systems into the condition $\delta_{n_{0}=0,k_{0}=0}^{+}=0$. Applying the selective interaction in Eq.~(\ref{HaTC}) to the initial state $\ket{D_{4}^{0}}\otimes \ket{0}$, one can obtain the superposition state $\ket{\psi(t)} = \cos\left(\Omega_{0,0}(\lambda)t\right)\ket{D_{4}^{0}}\otimes \ket{0} -i\sin\left(\Omega_{0,0}(\lambda)t\right)\ket{D_{4}^{1}}\otimes \ket{1}$. After a time period $T_{1}=t_{1}=\pi/(2\Omega_{0,0}(\lambda))$, we obtain the state $\ket{\psi(t_{1})}=\ket{D_{4}^{1}}\otimes \ket{1}$. When the qubit frequency is $\omega_{q}=-\omega_{r}+U/4$, we can see that the resonance peak appears in Fig. \ref{fig3}(a), and the average excitation number $\langle \hat{a}^{\dag}\hat{a}\rangle$ and $\langle \hat{N}_{q} \rangle$ reach the maximum $1$. The Fig. \ref{fig3}(b) shows a perfect Rabi oscillation can only be realized in the pre-selected subspace $\left\{\ket{D_{4}^{0}}\otimes\ket{0}, {\ket{D_{4}^{1}}}\otimes\ket{1}\right\}$, while the population of the other quantum Dicke states is nearly zero. In Fig. \ref{fig4}, we switch on the selective interaction by tuning the frequency of the two-level systems to satisfy the condition $\delta_{n_{0}=0,k_{0}=1}^{-}=0$. Applying selective interaction on the state $\ket{\psi(t_{1})}=\ket{D_{4}^{1}}\otimes \ket{1}$ for a time period $T_{2}=t_{2}-t_{1}=\pi/(2\Omega_{0,1}(\lambda))$, we obtain the state $\ket{\psi(t_{2})}=\ket{D_{4}^{2}}\otimes \ket{0}$. Then tuning $\omega_{q}$ to satisfy the condition $\delta_{n_{0}=0,k_{0}=2}^{+}=0$ and acting the selective anti-TC interaction for a time period $T_{3}=t_{3}-t_{2}=\pi/(2\Omega_{0,2}(\lambda))$, we can be obtain the state $\ket{\psi(t_{3})}=\ket{D_{4}^{3}}\otimes \ket{1}$. The dynamics for the selective interaction are shown in Fig. \ref{fig5}. Finally, to obtain the Dicke state $\ket{D_{4}^{4}}$, we should tune the frequency $\omega_{q}$ to satisfy the condition $\delta_{n_{0}=0,k_{0}=3}^{-}=0$, and act the selective TC interaction on the state $\ket{\psi(t_{3})}$ for a time period $T_{4}=t_{4}-t_{3}=\pi/(2\Omega_{0,3}(\lambda))$. The final state is $\ket{\psi(t_{4})}=\ket{D_{4}^{4}}\otimes \ket{0}$. Tracing out the photon state, we obtained the Dicke state $\ket{D_{4}^{4}}$. As can be seen from Fig. \ref{fig6}, when the parameters and resonance frequency are properly selected, the desired target state can be successfully prepared. In all of these pictures, we present the average excitation number and the population of Dicke states with the parameters $\omega_{r}=1$, $\lambda=0.006\omega_{r}$ and $U=-0.5\omega_{r}$.

\subsection{Generation GHZ state with selective TC and anti-TC model}
\begin{figure}[ht]
\centering
\includegraphics[angle=0,width=0.45\textwidth]{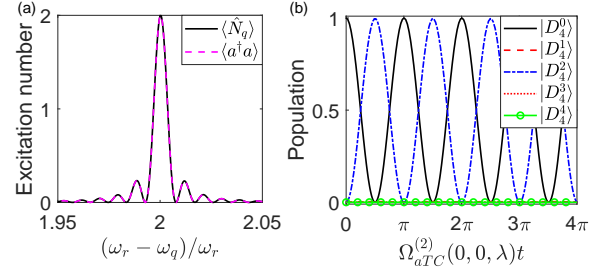}
\caption{Two-atom-excitation selective interactions of the Dicke-stark model with initial states $\ket{D_{4}^{0}}\otimes \ket{0}$. (a) After a time $t_{1}=\pi/(2\Omega_{\rm aTC}^{(2)}(0,0,\lambda))$, we calculation the average excitation of the particles for different ratios of $(\omega_{r}-\omega_{q})/\omega_{r}$, where the anti-TC peaks appears for $(\omega_{r}-\omega_{q})/\omega_{r}=2.0003$ which corresponds to $\tilde{\delta}_{\rm aTC}^{(2)}(0,0)=0$; (b)The population of Dicke states as a function of evolution time. The resonance frequency is chosen as $(\omega_{r}-\omega_{q})/\omega_{r}=2.0003$. The other parameters are given as follows: $\omega_{r}=1$, $\lambda/\omega_{r}=0.1$ and $U/ \omega_{r}=-16$.}
\label{fig7}
\end{figure}

The Greenberger-Horne-Zeilinger (GHZ) states, which can be viewed as superposition Dicke states, play very important role in several protocols of quantum communication and cryptography \cite{GHZ1989}. Here we will show how to generate GHZ state with selective two-atom excitation TC and anti-TC model. The $N$ qubits GHZ state reads
\begin{equation}
    \ket{GHZ}_{N}=\frac{1}{\sqrt{2}}\left(\ket{g_{1}g_{2}\cdots g_{N}}+e^{i\varphi}\ket{e_{1}e_{2}\cdots e_{N}}\right).
\end{equation}
Such state can be recast as superposition of Dicke states $\ket{GHZ}_{N}=\frac{1}{\sqrt{2}}\left(\ket{D_{N}^{0}}+e^{i\varphi}\ket{D_{N}^{N}}\right)$. We will show how to create four-qubit GHZ state with selective interaction. First, we prepare the initial state in $\ket{\psi(0)}=\ket{D_{4}^{0}}\otimes \ket{0}$ with $\omega_{r}=1$, $\lambda/\omega_{r}=0.1$ and $U/ \omega_{r}=-16$. In Fig. \ref{fig7}(a), we can find the frequency near the resonant point $\tilde{\delta}_{\rm aTC}^{(2)}(0,0)=0$, and when the resonance condition is satisfied, the excitation number is 2. Tuning the resonance condition $\tilde{\delta}_{\rm aTC}^{(2)}(0,0)=0$ and applying the selective two atoms anti-TC interaction in Eq.~(\ref{H2aTC}) on the initial state for a time period $t_{1}=\pi/(4\Omega_{\rm aTC}^{(2)}(0,0,\lambda))$. In Fig. \ref{fig7}(b), we obtain 
\begin{equation}
    \ket{\psi(t_{1})}=\frac{1}{\sqrt{2}}\left(\ket{D_{4}^{0}}\otimes \ket{0} -i\ket{D_{4}^{2}}\otimes \ket{2}\right)
\end{equation}

\begin{figure}[ht]
\centering
\includegraphics[angle=0,width=0.45\textwidth]{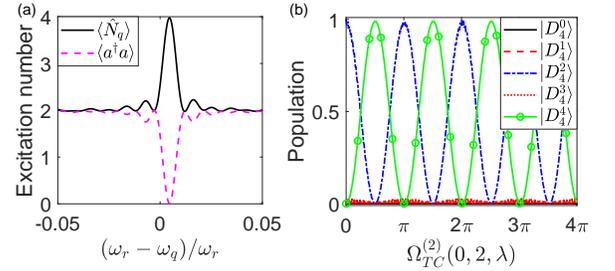}
\caption{Two-atom-excitation selective interactions of the Dicke-stark model. (a) After a time $t_{2}-t_{1}=\pi/(2\Omega_{\rm TC}^{(2)}(0,2,\lambda))$, we calculation the average excitation of the particles for different ratios of $(\omega_{r}-\omega_{q})/\omega_{r}$ and initial state $\ket{D_{4}^{2}}\otimes \ket{2}$, where the TC peaks appears for $(\omega_{r}-\omega_{q})/\omega_{r}=0.0046$ which corresponds to $\tilde{\delta}_{\rm TC}^{(2)}(0,2)=0$; (b)The population of Dicke states with initial state $\ket{D_{4}^{2}}$, and resonance frequency is selected as $(\omega_{r}-\omega_{q})/\omega_{r}=0.0046$. The other parameters are given as follows: $\omega_{r}=1$, $\lambda/\omega_{r}=0.1$ and $U/ \omega_{r}=-16$.}
\label{fig8}
\end{figure}

In Fig. \ref{fig7}(b), We can see that when the resonance frequency is satisfied, only the pre-selected subspace will oscillate periodically, while the other quantum states will not change. Then we tuning $\omega_{q}$ to satisfy the resonance condition $\tilde{\delta}_{\rm TC}^{(2)}(0,2)=0$ in Fig. \ref{fig8}(a), and applying the selective two atoms TC interaction in Eq.~(\ref{H2TC}) on the initial state. As can be seen from Fig. \ref{fig8}(b), the selective interaction evaluates the state $\ket{D_{4}^{2}}\otimes \ket{2}$ to superposition between $\ket{D_{4}^{2}}\otimes \ket{2}$ and $\ket{D_{4}^{4}}\otimes \ket{0}$, but remains the state $\ket{D_{4}^{0}}\otimes \ket{0}$. At time $t_{2}=t_{1}+\pi/(2\Omega_{\rm TC}^{(2)}(0,2,\lambda))$, the evolved state reads
\begin{equation}
    \ket{\psi(t_{2})}=\frac{1}{\sqrt{2}}\left(\ket{D_{4}^{0}} -\ket{D_{4}^{4}}\right)\otimes \ket{0}=\ket{GHZ}_{4}\otimes \ket{0}
\end{equation}
Then we obtained the four qubit GHZ state in two steps. To assess the performance of this proposal, we have simulated the final states via two steps numerically, and the final state denoted by $\ket{\psi_{f}(t_{2})}$. Then we compared the final state with the target state $\ket{\psi(t_{2})}$. The fidelity of these two states is $F=|\langle \psi_{f}(t_{2})\ket{\psi(t_{2})}|^{2}=0.9952$. To obtain higher fidelity of final state, more strongly nonlinear strength is needed. We also can generate other superposition of Dicke states by tuning the qubit frequency and controlling the evolution time. In Figs. \ref{fig7} and \ref{fig8}, we present the average excitation number and the population of Dicke states with the parameters $\omega_{r}=1$, $\lambda/\omega_{r}=0.1$ and $U/\omega_{r}=-16$.

\section{Conclusion}
To summarize, we have studied the selective TC and anti-TC interaction can be engineered by means of Dicke model with Stark term. In particular, we show the selective interaction of one atomic and two atomic excitations. Their resonance conditions and oscillation frequencies depend on the number of excited atoms and the number of excited photons. In the pre-selected subspace, we can choose appropriate parameters to create selective interactions. Then we can achieve the preparation of the target state, the main process is accomplished according to RWA. With the aid of numerical simulations, we verified the correctness of the theory and successfully prepared Dicke states with very high fidelity and GHZ states which using high-order effective Hamiltonian.

\section*{Acknowledgements}
The work is supported by National Natural Science Foundation of China (Grant No. 201118183), Fundamental Research Funds for the Central Universities (Grant No. 2412020FZ026) and Natural Science Foundation of Jilin Province (Grant No. JJKH20190279KJ).

\bibliography{ref}

\end{document}